\documentclass[prl,amsmath,amssymb,aps,twocolumn,superscriptaddress,showpacs
]{revtex4-1}
\usepackage{graphicx}
\usepackage{dcolumn}
\usepackage{bm}

\begin{document}

%\preprint{}

\title{Coherent transfer of photoassociated molecules into the rovibrational ground state}


\author{K. Aikawa}
\affiliation{Department of Applied Physics, The University of Tokyo, Hongo, Bunkyo-ku, Tokyo 113-8656, Japan}
\email{ka\_cypridina@atomtrap.t.u-tokyo.ac.jp}
\author{D. Akamatsu}
\affiliation{Institute of Engineering Innovation, The University of Tokyo, Yayoi, Bunkyo-ku, Tokyo 113-8656, Japan}
\author{M. Hayashi}
\affiliation{Department of Applied Physics, The University of Tokyo, Hongo Bunkyo-ku, Tokyo 113-8656, Japan}
\author{K. Oasa}
\affiliation{Department of Applied Physics, The University of Tokyo, Hongo, Bunkyo-ku, Tokyo 113-8656, Japan}
\author{J. Kobayashi}
\affiliation{Institute of Engineering Innovation, The University of Tokyo, Yayoi, Bunkyo-ku, Tokyo 113-8656, Japan}
\author{P. Naidon}
\affiliation{JST, ERATO, Yayoi, Bunkyo-ku, Tokyo 113-8656, Japan}
\author{T. Kishimoto}
\affiliation{Center for Frontier Science and Engineering, University of Electro-communications, Chofu, Tokyo 182-8585, Japan}
\author{M. Ueda}
\affiliation{JST, ERATO, Yayoi, Bunkyo-ku, Tokyo 113-8656, Japan}
\affiliation{Department of Physics, The University of Tokyo, Yayoi, Bunkyo-ku, Tokyo 113-8656, Japan}
\author{S. Inouye}
\affiliation{Institute of Engineering Innovation, The University of Tokyo, Yayoi, Bunkyo-ku, Tokyo 113-8656, Japan}
\affiliation{JST, ERATO, Yayoi, Bunkyo-ku, Tokyo 113-8656, Japan}
%\affiliation
%\thanks{}
%\affiliation{}


\date{\today}

\begin{abstract}
We report on the direct conversion of laser-cooled $^{41}$K and $^{87}$Rb atoms into ultracold $^{41}$K$^{87}$Rb molecules 
in the rovibrational ground state via photoassociation followed by stimulated Raman adiabatic passage. 
High-resolution spectroscopy based on the coherent transfer revealed the hyperfine structure of weakly bound molecules in an unexplored region. 
Our results show that a rovibrationally pure sample of ultracold ground-state molecules is achieved via the all-optical association of laser-cooled atoms, 
opening possibilities to coherently manipulate a wide variety of molecules. 
\end{abstract}

%\keywords{}

\maketitle

Progress in cooling and manipulating atoms has led to diverse applications. Even more possibilities will open up
 if molecules are cooled and prepared in the rovibrational ground state because then the translational, vibrational, and 
rotational degrees of freedom of the molecules can be manipulated with greater accuracy.
The potential applications range from novel quantum phases to ultracold chemistry, quantum computation, 
and precision measurements \cite{krems2009cold, *carr2009cold}. Various schemes to achieve cold molecules have been 
demonstrated thus far. Direct cooling of molecules has been limited in terms of temperature 
\cite{weinstein1998magnetic, *bethlem1999decelerating, *elioff2003subkelvin, *gupta1999mechanical, *rangwala2003continuous}. 
Recently, quantum gases of molecules in the rovibrational ground state have been achieved 
\cite{ni2008high, lang2008ultracold, danzl2010ultracold} with a coherent optical transfer of weakly bound molecules 
produced via the magnetoassociation \cite{kohler2006production} of quantum degenerate atoms. However, 
this scheme is only applicable to atomic species for which magnetoassociation is available. 
An alternative general method for producing weakly bound molecules from ultracold atoms is photoassociation (PA) \cite{jones2006ultracold}. 
Up to now, only the incoherent formation of molecules in the vibrational ground state via photoassociation 
has been reported \cite{sage2005optical, viteau2008optical, deiglmayr2008formation}. 
Although the direct photoassociation yielded molecules in the $v''=0$, $J''=2$ level predominantly 
\cite{deiglmayr2008formation}, the state-selective photoassociative formation of cold molecules in 
the rovibrational ground state is yet to be achieved.

\begin{figure}
\includegraphics[width=75mm]{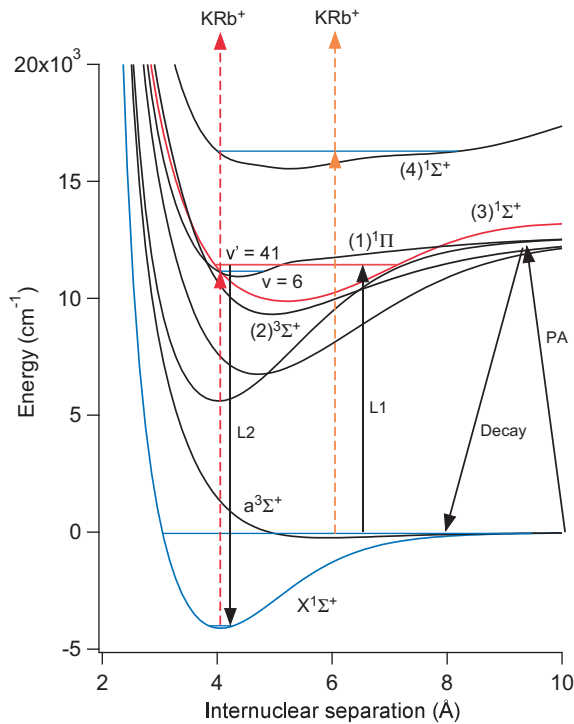}
\caption{\label{fig:KRbPEC}Relevant molecular potential energy curves of KRb. Weakly bound molecules were produced by the 
photoassociation of laser-cooled $^{41}$K and $^{87}$Rb atoms. Molecules in the $v''=91$, $J''=0$, $F''=2$ level of $X^{1}\Sigma^{+}$ 
were coupled to the $v''=0$, $J''=0$ level of $X^{1}\Sigma^{+}$ by a two-photon transition with wavelengths near 875(L1) and 641(L2) nm 
via the $v'=41$, $J'=1$ level of $(3)^{1}\Sigma^{+}$. For depletion spectroscopy and SpIDR spectroscopy, 
two-photon ionization through the $(4)^{1}\Sigma^{+}$ state was used. For the detection of the rovibrational ground-state molecules, 
a three-photon transition mediated by the $v'=6$ level of $(1)^{1}\Pi$ was used.
}
\end{figure}

Here, we report that a rotationally and vibrationally pure sample of ultracold $^{41}$K$^{87}$Rb molecules 
in the rovibrational ground state could be directly produced in a magneto-optical trap (MOT) of $^{41}$K 
and $^{87}$Rb via photoassociation followed by stimulated Raman adiabatic passage (STIRAP) 
\cite{bergmann1998coherent, *vitanov2001coherent, *kral2007colloquium}. In spite of the low laser 
intensity due to the large volume of our molecular gas, we could reach a STIRAP transfer efficiency of more than 
70\% by employing an intermediate state with a narrow natural linewidth. 
In order to identify the two-photon resonance precisely, we developed a new detection scheme, 
spontaneous-decay-induced double resonance (SpIDR), which provided the high-resolution 
spectra of one-photon transitions from PA molecules and allowed us to observe 
a two-photon dark resonance \cite{harris97eit, *lukin2003colloquium, *fleischhauer2005electromagnetically}. 
Our results show that a rovibrationally pure sample of ultracold molecules in the ground state is achieved 
via the all-optical association of laser-cooled atoms, opening novel possibilities to coherently manipulate a wide variety of cold molecules 
that can be produced via photoassociation. 

Fig.~\ref{fig:KRbPEC} shows the schematic representation of the coherent transfer of ultracold $^{41}$K$^{87}$Rb 
molecules. The potential curves are taken from Ref.~\onlinecite{rousseau2000theoretical}. 
Details of the experimental setup for the production and detection of weakly bound molecules is found elsewhere \cite{aikawa2009toward}. 
For the present study, we used $^{41}$K$^{87}$Rb molecules in the $v''=91$, $J''=0$ level of $X^{1}\Sigma^{+}$ produced via photoassociation
of $^{41}$K and $^{87}$Rb atoms in a MOT. The typical numbers, densities and temperatures of atoms were $1\times 10^{8}$, 
$2\times 10^{11}$ cm$^{-3}$ and 400 $\mu$K for $^{41}$K and $2\times 10^{8}$, $4\times 10^{11}$ cm$^{-3}$ and 100 $\mu$K for $^{87}$Rb, respectively. 
The PA laser resonant to the $J'=1$ level of $\Omega=0^{+}$ 
(wave number $\sim12570$ cm$^{-1}$, intensity $\sim1\times10^{3}$ W cm$^{-2}$) was applied for 10 ms. 
Molecules were ionized through resonance enhanced multi-photon ionization (REMPI) by 
a pulsed dye laser and detected by micro-channel plates (MCP). The entire experimental procedure was operated at 9 Hz.

Despite the successful demonstrations on quantum gases, the STIRAP transfer into 
the rovibrational ground state was not straightforward for PA molecules. 
Any residual magnetic field could cause decoherence for Raman coupling between two molecular states 
since our molecular sample was not spin-polarized. In addition, untrapped, high-temperature 
PA molecules occupied a volume larger ($\sim1$ ${\rm mm}^{3}$) than that occupied by the trapped quantum gases ($\sim10^{-4}$ ${\rm mm}^{3}$). 
This limited the available laser intensity for driving the transition.

These difficulties were circumvented by starting from singlet molecules which was insensitive to magnetic fields. Utilizing singlet 
excited states with a narrow natural linewidth as an intermediate state, we realized an efficient 
Raman transfer of PA molecules into a singlet ground state. This was in contrast with previous works 
on the two-photon transfer of heteronuclear molecules \cite{ni2008high, sage2005optical}, 
which started from triplet molecules and transferred the PA molecules into the singlet ground state 
via the $\Omega=1$ state (the spin-mixed state of $(2)^{3}\Sigma^{+}$, $(1)^{3}\Pi$ and $(1)^{1}\Pi$). A transition strength comparable to the $\Omega=1$ 
state was obtained with the $v'=41$ level of $(3)^{1}\Sigma^{+}$ \cite{wang2007rotationally, aikawa2009toward}, 
whereas the narrow natural linewidth of this state ($\Gamma\sim2\pi\times$300 kHz, calculated from Ref.~\onlinecite{beuc2006predictions}) 
allowed us to obtain an efficient transfer with a relatively low laser intensity. 

Our optical system for the Raman transition consisted of four lasers. Two diode lasers (875 nm and 641 nm) were locked to an 
ultra-low expansion (ULE) cavity with a dual-wavelength coating. These lasers worked as master lasers. 
Two slave lasers, a ring dye laser (641 nm) and a diode laser (875 nm), were then locked to 
master lasers through an optical phase-locked loop (OPLL) \cite{ricci1995compact} and used for experiments. 
The scanning of the Raman lasers was realized by sweeping the microwave source for OPLL. 
In this report, the microwave frequency for OPLL was noted as an offset frequency. 
The typical short-term linewidth of the Raman lasers was less than 10 Hz, whereas the long-term 
stability of the ULE cavity was estimated to be less than 100 kHz by monitoring the Cs D2 transition. 
The $1/e^{2}$ diameter of the Raman lasers was set as 1.5 mm which was larger than the size of atomic clouds. 
For the STIRAP transfer, the power of the laser L1 and the laser L2 was set to 9 mW and 25 mW, respectively. 

\begin{figure}
\includegraphics[width=80mm]{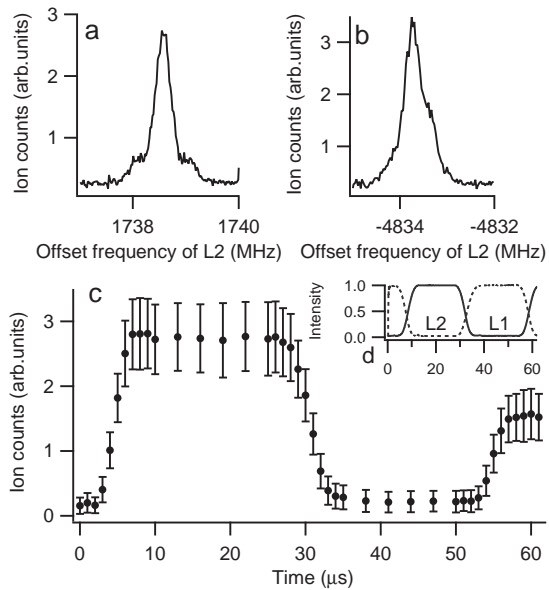}
\caption{\label{fig:STIRAP}STIRAP transfer from the $v''=91$, $J''=0$, $F''=2$ level to the $v''=0$ level of $X^{1}\Sigma^{+}$. 
(a) Transfer to the $J''=0$ level. The population in the $v''=0$ level of $X^{1}\Sigma^{+}$ is plotted against the frequency 
of the down transition laser (L2). (b) Transfer to the $J''=2$ level. The asymmetry observed in the $J''=2$ 
spectrum is expected to be a result of the hyperfine structure induced by the nuclear electric 
quadrupole moments of $^{87}$Rb. (c) Time evolution of the population in the $v''=0$, $J''=0$ level of $X^{1}\Sigma^{+}$ 
during the multiple STIRAP transfer process. The error bars indicate the standard deviation. 
(d) Time variation of the intensities of the Raman lasers during the multiple-transfer process. 
Both intensities are normalized to unity. During the waiting time of 20 $\mu$s set 
between the transfers, the remaining population that is not transferred by STIRAP is completely removed by the resonant Raman lasers. }
\end{figure}

Fig.~\ref{fig:STIRAP}a shows the population in the $v''=0$ level of $X^{1}\Sigma^{+}$ after the STIRAP transfer into the $J''=0$ level as a function of 
the frequency of the down transition laser (L2). The observed two-photon linewidth of 200 kHz was 
consistent with a simple three-level theoretical model \cite{bergmann1998coherent, *vitanov2001coherent, *kral2007colloquium} 
when we took into account the Rabi frequency ($2\pi\times$400 kHz) and the transfer duration (10 $\mu$s). 
In order to confirm that Fig.~\ref{fig:STIRAP}a indicated the rovibrational ground state, we also performed a 
STIRAP transfer into the $v''=0$, $J''=2$ level (Fig.~\ref{fig:STIRAP}b) and determined the rotational 
constant to be 1095.4(1) MHz$\times h$  ($h$ is the Planck's constant). The obtained value was in good agreement 
with the calculated value of 1095.362(5) MHz$\times h$ obtained from the mass-scaling of the experimental value for $^{40}$K$^{87}$Rb \cite{ospelkaus2010controlling}. 
The typical number, density and temperature of the produced ground-state molecules were $10^{3}$, $10^{6}$ ${\rm cm}^{-3}$, and 130 $\mu K$, respectively. 
The production rate in the rovibrational ground state was as high as $10^{4}$ ${\rm s}^{-1}$.

The direct evidence of the STIRAP transfer was obtained by studying the time evolution of the population 
in the ground state (Fig.~\ref{fig:STIRAP}c) during the multiple transfer process (Fig.~\ref{fig:STIRAP}d). The ground-state molecules 
after a single transfer were used for measuring the efficiency of transfer. We typically retrieved 53\% 
of the ground-state molecules after the double-transfer process, which implied a single-step efficiency of 73\%. 
Currently, the efficiency was limited by the difference in the Doppler shift for the two lasers.

\begin{figure}
\includegraphics[width=80mm]{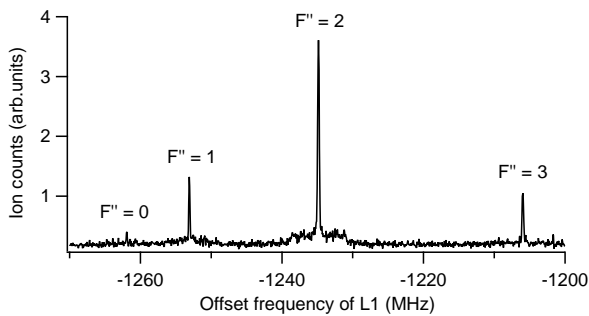}
\caption{\label{fig:STIRAPspectroscopy}STIRAP spectroscopy of weakly bound molecules (the $v''=91$, $J''=0$ level of $X^{1}\Sigma^{+}$). 
An efficient transfer into the rovibrational ground state enables the precision spectroscopy of the hyperfine structure 
of photoassociated molecules in an unexplored region with a typical accuracy of 10 kHz. 
The labeling is based on the total angular momentum including the nuclear spin.}
\end{figure}

The STIRAP transfer into the rovibrational ground state can be used as new precision spectroscopy for 
weakly bound molecules in an unexplored region (Fig.~\ref{fig:STIRAPspectroscopy}). By mapping the population in the initial 
hyperfine levels onto the ground state, we could measure the hyperfine splitting of PA molecules with an accuracy of 10 kHz$\times h$. 
A higher resolution could be achieved by using a longer transfer duration and smaller intensities for the Raman lasers. 
The observed energy levels of the $F''=3, 2, 1$ levels with respect to the $F''=0$ level were 55.960(9), 27.089(7), 
and 8.852(7) MHz$\times h$, respectively. The values of these energy levels as obtained by a coupled channel 
calculation were 56.180, 27.176, and 8.870 MHz$\times h$; these values were in reasonable agreement with our observations. 

The rest of this report is devoted to SpIDR spectroscopy, which is an essential improvement of the 
one-photon spectroscopy of PA molecules. This improvement was necessary for identifying the 
structure of the excited states and observing a two-photon dark resonance. To date, depletion spectroscopy 
has been widely used for analyzing the transition from PA molecules \cite{wang2007rotationally}. A continuous-wave (CW) laser 
depletes the number of PA molecules, while the loss is monitored with REMPI. Although this spectroscopy was 
straightforward, we encountered a serious problem related to the signal-to-noise ratio when we attempted to 
record data at a low intensity in order to avoid saturation broadening. At a relatively high intensity, 
where we could observe a large signal, the spectral width broadened to tens of megahertz, 
washing out all the structure from the hyperfine levels. In addition, saturation broadening smoothed out 
the transparency peak of the two-photon dark resonance and made it difficult to identify the two-photon resonance.

In order to improve the signal-to-noise ratio of depletion spectroscopy, we developed a method to 
monitor not the depletion but the creation of molecules in the excited state. In practice, 
we monitored the population in one of the vibrational levels in the ground state to which a significant fraction 
of excited molecules decay. Thus, we could obtain a background-free spectrum, thereby increasing 
the signal-to-noise ratio to observe the narrow linewidth of the $(3)^{1}\Sigma^{+}$ state. 

\begin{figure}
\includegraphics[width=80mm]{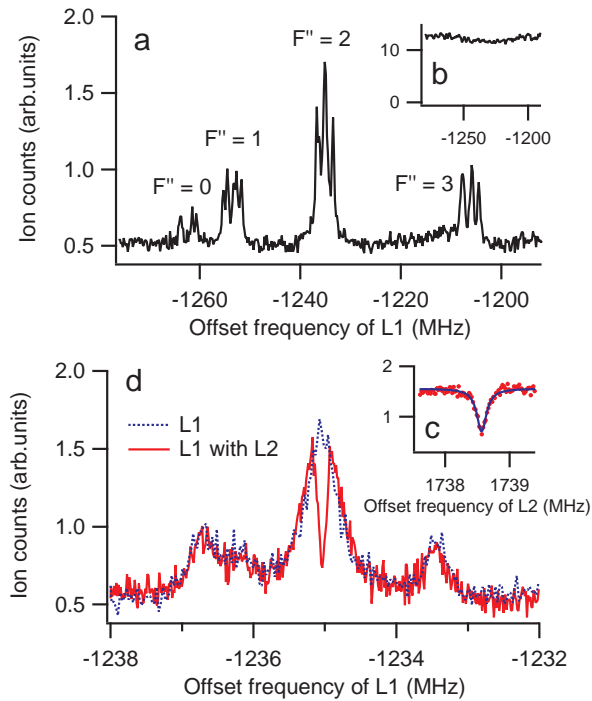}
\caption{\label{fig:Darkresonance}SpIDR spectroscopy of photoassociated molecules. 
(a) SpIDR spectrum of the $v'=41$, $J'=1$ level of $(3)^{1}\Sigma^{+}$ taken by exciting molecules in the $v''=91$, 
$J''=0$ level of $X^{1}\Sigma^{+}$. Molecules formed by spontaneous decays from the excited state are counted 
against the offset frequency of the laser L1. The coarse splitting results from the hyperfine 
structure of the $v''=91$, $J''=0$ level of $X^{1}\Sigma^{+}$. The fine splitting in each line implies the existence of 
hyperfine structure in the excited state. For the two-photon transition, the largest peak ($F''=2$) is 
employed as an initial state. (b) Depletion spectrum. In order to observe any signal, we had to 
increase the laser intensity by three orders of magnitude as compared to (a). Transitions from the $v''=91$, 
$J''=0$ level are barely observable since the ion counts are dominated by molecules in the $v''=91$, $J''=2$ level. 
(c) Two-photon dark resonance between the $v''=91$, $J''=0$, $F''=2$ level and the $v''=0$, $J''=0$ level of $X^{1}\Sigma^{+}$. 
The down transition laser (L2) is scanned while the up transition laser (L1) is held on resonance. 
The dots indicate experimentally obtained data and the solid curve is a fit to the data points. 
(d) Two-photon dark resonance obtained by scanning the laser L1. The laser L2 is held on resonance. 
The blue dashed curve indicates a SpIDR spectrum for the $F''=2$ level, while the red solid one indicates a spectrum obtained with the laser L2.}
\end{figure}

Fig.~\ref{fig:Darkresonance}a shows the SpIDR spectrum of the $v'=41$, $J'=1$ level of $(3)^{1}\Sigma^{+}$. A considerably smaller power (5 $\mu$W) 
and shorter time (400 $\mu$s) than those required for depletion spectroscopy ($>100$ $\mu$W, 10 ms) were sufficient to 
detect the excitation. This led to a significantly narrow linewidth close to the natural linewidth. 
The coarse splitting resulted from the hyperfine structure in the ground state, whereas the fine splitting 
implied the existence of the hyperfine structure in the excited state. For the sake of comparison, 
the depletion spectrum in the same frequency range is shown in Fig.~\ref{fig:Darkresonance}b. 

The SpIDR spectroscopy enabled us to observe a two-photon dark resonance (Fig.~\ref{fig:Darkresonance}c). 
The dip indicated the emergence of the two-photon dark state with the laser L2 \cite{harris97eit, *lukin2003colloquium, *fleischhauer2005electromagnetically}. 
By comparing the two frequencies at the dip, we determined the binding energy of the $v''=0$, $J''=0$ level of $X^{1}\Sigma^{+}$ with respect to 
the $v''=91$, $J''=0$ level of $X^{1}\Sigma^{+}$ to be -124955.92(4) GHz$\times h$. Taking into account the binding energy of the $v''=91$, $J''=0$ level 
with respect to the threshold $F_{\rm K}=1+F_{\rm Rb}=1$, -374.75(3) GHz$\times h$, the binding energy of the $v''=0$, $J''=0$ level 
with respect to the atomic level without the hyperfine structure was determined to be -125335.11(5) GHz$\times h$. 
From the width of the transparency peak, we determined the transition dipole moment for the down transition to be 0.0098(7) $ea_0$, 
assuming the natural linewidth of the excited state to be $2\pi\times$300 kHz. When we locked the laser L2 to the dip frequency 
and scanned the laser L1, we obtained a well-known dark resonance spectrum (Fig.~\ref{fig:Darkresonance}d). 

In conclusion, we realized an efficient STIRAP transfer of PA molecules into the rovibrational ground state. 
High-resolution spectroscopy of the hyperfine structure of PA molecules in an unexplored region was demonstrated using coherent transfer. 
We developed a highly sensitive detection scheme which enabled us to take a high-resolution spectrum of 
one-photon transitions and observe a two-photon dark resonance for PA molecules. 
Our method is readily extended to other atomic species for which laser cooling and photoassociation is available. 
Trapping molecules into an optical dipole trap \cite{hudson2008inelastic} is the next key issue. 
The trapped sample of molecules will constitute a test tube for ultracold chemistry. Moreover, the high repetition rate 
makes our method a promising candidate for high-precision spectroscopy. The precision measurement of 
the electron-to-proton mass ratio in cold molecules \cite{zelevinsky2008precision, *demille2008enhanced} is within reach with our methods.

\begin{acknowledgments}
We thank M. Kozuma for helpful discussions, and M. Falkenau, Y. Tanooka, and K. Mori for experimental assistance 
in the early stages of the experiment. K. A. and D. A. acknowledge support from the Japan Society for the Promotion of Science.
\end{acknowledgments}

\providecommand{\noopsort}[1]{}\providecommand{\singleletter}[1]{#1}%

\end{document}